\documentclass[final]{raa06}           %  For uploading to arXiv and final 
\usepackage{graphicx,times}             %for PS/EPS graphics inclusion, new
\usepackage{natbib}
\usepackage{amssymb,amsmath}
\bibpunct{(}{)}{;}{a}{}{,}
\usepackage[colorlinks=true, citecolor=blue]{hyperref}%
\usepackage{graphicx,kantlipsum,setspace}
\usepackage{caption}
\usepackage{graphics,epsf}
\usepackage{amsmath}                % American Mathematical Society package
\usepackage{amsfonts}               % American Mathematical Society fonts
\usepackage{amssymb}                % American Mathematical Society symbol
\usepackage{epsfig}                 % EPS figures
\usepackage{appendix}
\usepackage{graphicx}
\usepackage{float}
\usepackage{color}
\usepackage{multirow}
\usepackage{colortbl}
\usepackage[para,online,flushleft]{threeparttable}
\usepackage{xcolor}

\hypersetup{citecolor=blue, % color for \cite{...} links
            linkcolor=red, % color for \ref{...} links
            menucolor=blue, % color for Acrobat menu buttons
            urlcolor=blue}  % color for \url{...} links
\newcommand{\cm}{{~\rm cm}}

\newcommand{\km}{{~\rm km}}
\newcommand{\s}{{~\rm s}}

\newcommand{\yr}{{~\rm yr}}

\newcommand{\keV}{{~\rm keV}}

\begin{document}

   \title{Classifying core collapse supernova remnants by their morphology as shaped by the last exploding jets
%\footnotetext{$*$ Supported by the National Natural Science Foundation of China.}
}
%   \subtitle{I. Place Your Subtitle Here}

   \volnopage{Vol.0 (20xx) No.0, 000--000}      %%preserved for Editor. DOn't remove!
   \setcounter{page}{1}          %%starting page, preserved for Editor. DOn't remove!

   \author{Noam Soker
     % \inst{1}
    }

   \institute{Department of Physics, Technion, Haifa, 3200003, Israel;   {\it   soker@physics.technion.ac.il}\\
%        \and
%             Full institute address for the third author\\
\vs\no
   {\small Received~~20xx month day; accepted~~20xx~~month day}}

\abstract{Under the assumption that jets explode all core collapse supernovae (CCSNe) I classify 14 CCSN remnants (CCSNRs) into five groups according to their morphology as shaped by jets, and attribute the classes to the specific angular momentum of the pre-collapse core. \textit{Point-symmetry} (1 CCSNR): According to the jittering jets explosion mechanism (JJEM) when the pre-collapse core rotates very slowly the newly born neutron star (NS) launches tens of jet-pairs in all directions. The last several jet-pairs might leave an imprint of several pairs of `ears’, i.e., a point-symmetric morphology.  \textit{One pair of ears} (8 CCSNRs): More rapidly rotating cores might force the last pair of jets to be long-lived and shape one pair of jet-inflated ears that dominate the morphology. \textit{S-shaped} (1 CCSNR): The accretion disk might precess, leading to an S-shaped morphology. \textit{Barrel-shaped} (3 CCSNRs): Even more rapidly rotating pre-collapse cores might result in a final energetic pair of jets that clear the region along the axis of the pre-collapse core rotation and form a barrel-shaped morphology. \textit{Elongated} (1 CCSNR): Very rapidly rotating pre-collapse core force all jets to be along the same axis such that the jets are inefficient in expelling mass from the equatorial plane and the long-lasting accretion process turns the NS into a black hole (BH). The two new results of this study are the classification of CCSNRs into five classes based on jet-shaped morphological features, and the attribution of the morphological classes mainly to the pre-collapse core rotation in the frame of the JJEM. 
}

 \authorrunning{N. Soker}            
\titlerunning{Classifying CCSN remnants by last exploding jets}  
   
      \maketitle

% =========================================
\section{Introduction} 
\label{sec:intro}
% =========================================

There is no consensus on the explosion mechanism of core collapse supernovae (CCSNe). There are two competing theoretical explosion mechanisms that are based on the gravitational energy that the formation process of the newly born neutron star (NS) or black hole (BH) releases as the core of the CCSN progenitor collapses. These mechanisms are the delayed neutrino explosion mechanism (\citealt{BetheWilson1985}, followed by hundreds of studies since then, e.g., \citealt{Hegeretal2003, Janka2012, Nordhausetal2012, Mulleretal2019Jittering, BurrowsVartanyan2021, Fujibayashietal2021, Fryeretal2022, Bocciolietal2022, Nakamuraetal2022, Olejaketal2022}), and the jittering jets explosion mechanism (JJEM; \citealt{Soker2010}, with limited number of studies that followed \citealt{PapishSoker2011, GilkisSoker2015, Quataertetal2019, Soker2020RAA, ShishkinSoker2021, AntoniQuataert2022, Soker2022SNR0540, AntoniQuataert2023, Soker2023gap}). 

According to the JJEM intermittent accretion disks (or belts; e.g., \citealt{SchreierSoker2016}) with stochastically varying angular momentum axes launch pairs of jets that explode the star. Pre-collapse stochastic core convection motion (e.g., \citealt{Soker2010, PapishSoker2014Planar, GilkisSoker2015, Soker2019SASI, ShishkinSoker2022, Soker2022SNR0540, Soker2022Boosting}; in some cases envelope convection motion can supply these seed perturbations, e.g., \citealt{Quataertetal2019, AntoniQuataert2022, AntoniQuataert2023}) serve as seed angular momentum perturbations. Instabilities between the newly born NS and the stalled shock at $\simeq 100 \km$ from the NS amplify these seed perturbations to sufficiently large specific angular momentum fluctuations as to form the intermittent accretion disks (e.g., \citealt{ShishkinSoker2021}). In case of core rotation the stochastic angular momentum variations are around the angular momentum axis of the pre-collapse core (e.g., \citealt{Soker2023gap}).

There are some fundamental differences between the JJEM and many papers that study jet-driven explosions that operate only for rapidly rotating pre-collapse cores and therefore the jets that the newly born NS or BH launch have a fixed axis (e.g., \citealt{Khokhlovetal1999, Aloyetal2000, MacFadyenetal2001, Maedaetal2012, LopezCamaraetal2013, BrombergTchekhovskoy2016,  Nishimuraetal2017, WangWangDai2019RAA, Grimmettetal2021, Perleyetal2021, Gottliebetal2022, ObergaulingerReichert2023, Urrutiaetal2023a}). These differences are as follows (e.g., \citealt{Soker2022Rev}). (1) As explained above, the JJEM operates even when the pre-collapse core does not rotate. (2) The JJEM asserts that jets explode most, and possibly all, CCSNe. (3) This implies that there are no failed CCSNe in the frame of the JJEM. All massive stars explode, even when a BH is formed. 
(4) The JJEM operates in a jet negative feedback mechanism. Namely, when the jets manage to explode the star accretion stops (with some delay time). This accounts for explosion energies that are several times the binding energy of the ejected mass. 

There might be $\approx {\rm few} - 30$ jet-launching episodes during the entire explosion process with the following properties \citep{PapishSoker2014a}. The jets launching velocities are $\simeq 10^5 \km \s^{-1}$ (neutrino observations limit the jets in most cases to be non-relativistic, e.g. \citealt{Guettaetal2020}). 
The explosion time might be $\simeq 1 - 10 \s$, where each individual jet-launching episode lasts for $\simeq 0.01-0.1 \sec$, beside probably the last jet-launching episode that might in some cases be much longer, as I propose in this study. The two jets in each jet-launching episode carry a mass of $\approx 10^{-3} M_\odot$. During the explosion process the newly born NS accretes a mass of $\approx 0.1 M_\odot$ through intermittent accretion disks, i.e., each accretion disk of an episode has a mass of $\approx 10^{-2} M_\odot$. 
These properties can vary a lot from one CCSN to another because they depend on the convection motion in the pre-collapse core, its angular momentum, and the binding energy of the ejecta. 
 
As far as the basic outcomes of the explosions, e.g., nucleosynthesis and lightcurves, the JJEM is similar to the neutrino driven-mechanism. The JJEM includes also heating by neutrinos as a boosting process \citep{Soker2022Boosting}. The differences include the morphology of the ejecta and that the JJEM can explain also very energetic CCSNe. 
This study deals with the morphology that the late jets imprint on the ejecta.
Early jets are choked inside the core, deposit their energy in the core, and explode it. Instabilities in the JJEM develop similar, but not identical, to those in the neutrino-driven explosion mechanism (for the later see, e.g.,  \citealt{Wongwathanaratetal2015, Wongwathanaratetal2017, BurrowsVartanyan2021, Vartanyanetal2022}). The jets are expected to introduce a point-symmetrical morphological component to the instabilities and mixing of isotopes. By point-symmetry I refer to a structure where to each structural feature there is a counterpart on the other side of the center. Because of the highly-non-spherical explosion process the counter structural feature can have a different small-scale structure, can have a different brightness, and be at a different distance from the center. The best example is the supernova remnant (SNR) 0540-69.3 that I study in section \ref{subsubsec:SNR0540693} and which possesses point-symmetry in its inner regions \citep{Soker2022SNR0540}.  

In this study, however, I focus on late jets, namely, jets that the newly born NS or BH launch after the earlier jets exploded the core. 
I examine the morphological features that such jets imprint on the outer regions of the ejecta as observed in CCSN remnants (CCSNRs). 
 In section \ref{sec:SNRS} I classify 14 SNRs into five classes. 
In section \ref{sec:CoreRotation} I suggest that the main, but not sole, property that determines the class of a SNR is the pre-collapse core angular momentum. This proposed explanation, and actually this entire paper, is largely motivated by my recent proposed explanation for the NS to BH mass gap in the frame of the JJEM \citep{Soker2023gap}. 
I summarize this study in section \ref{sec:Summary}.

%% ==========================================
\section{Classification of SNRs} 
\label{sec:SNRS}
% ==========================================

I classify 14 CCSNRs into five classes. Many other CCSNRs morphologies are too `messy'  and do not allow classification into one of these classes, e.g., VRO 42.05.01 (G166.0+4.3; for an image see, e.g., \citealt{Xiaoetal2022}).  I describe each class in a separate subsection and in the same order as the classes appear in Table \ref{table:SNRs}. The first row of Table \ref{table:SNRs} lists the five classes and the lower rows lists the CCSNRs in each class. The second row refers to my suggestion as to the main (but not sole) effect that determines the morphological properties of the last jets to be launched in the explosion process according to the JJEM (section \ref{sec:CoreRotation}). I assume that the main shaping of the morphology is by jets and not by other processes, such as the magnetic field of the interstellar medium (e.g., \citealt{Wuetal2019, Velazquezetal2023}). { The variable $j_{\rm p}$ is the pre-collapse average specific angular momentum of the core material that the newly born NS accretes as it launches jets;`p' stands for pre-collapse rotation which has a fixed direction. The variable $j_{\rm f}$ is the amplitude of the fluctuations in the specific angular momentum of the material that the NS accretes due to the velocity fluctuations of the pre-collapse convective zone. The amplitude is after instabilities amplify the perturbations. The direction of this angular momentum component varies stochastically; `f' stands for fluctuating directions. }
% TTTTTTTTTTTTTTTTTTTTTTTTTTTTTTTTTTTTTTTTTTTTTTT
\begin{table*}[]%[t]
        \centering
 \begin{tabular}{|p{2.0cm}|p{5.5cm}|p{2.9cm}|p{2.5cm}|p{1.3cm}|}
%    \begin{tabular}{|c|c|c|c|c|c|}
\hline
Point-Symmetry & One pair of ears& S-shaped & Barrel-shaped  & Elongated  \\ 
\hline   
$j_{\rm p} \lesssim 0.01 j_{\rm f}$&   $0.01j_{\rm f} \lesssim j_{\rm p} \lesssim 0.1 j_{\rm f}$  & $0.01 j_{\rm f} 
 \lesssim j_{\rm p} \lesssim 0.1 j_{\rm f}$   &  $j_{\rm p} \approx 0.1 j_{\rm f}-0.3 j_{\rm f}$   &   $j_{\rm p} \gtrsim j_{\rm f}$   \\ 
\hline
  Vela$^{[\ref{Fig:Vela1}]}$ & 0540-69.3$^{[\ref{fig:SNR0540693}]}$; Cassiopeia A$^{[\ref{Fig:CassiopeiaA}]}$; 3C58$^{[\ref{Fig:CassiopeiaA}]}$;  &   W44$^{[\ref{Fig:W44}]}$ &   RCW 103$^{[\ref{Fig:RCW103}]}$ &  W50$^{[\ref{Fig:W50}]}$    \\ 
(0540-69.3)$^{\#}$  &  S147$^{[\ref{Fig:CassiopeiaA}]}$; G290.1-0.8$^{[\ref{Fig:G290}]}$; & & G292.0+1.8$^{[\ref{Fig:G292.0+1.8}]}$    &      \\ 
& N49B$^{[\ref{Fig:N49B}]}$; Puppis A$^{[\ref{Fig:N49B}]}$; Crab Nebula$^{[\ref{Fig:N49B}]}$ & & G309.2-00.6$^{[\ref{Fig:G3092006}]}$ & \\
\hline
    \end{tabular}
    \caption{The classification of CCSNRs into five classes according to the last jets to be launched in the explosion. The second row lists the relation between the pre-collapse average specific angular momentum of the core  $j_{\rm p}$, and the magnitude of the stochastic fluctuations in the specific angular momentum of the mass that the newly born NS or BH accrete, $j_{\rm f}$ (see section \ref{sec:CoreRotation}). Comments: \# The inner structure of SNR 0540-69.3 is point symmetric. However, in this study I focus on the last jets to be launched, and therefore I include this SNR in the one-pair class (Fig. \ref{fig:SNR0540693}). Small numbers inside square parentheses are the figures where I present the CCSNRs.    
    }
      \label{table:SNRs}
\end{table*}
% TTTTTTTTTTTTTTTTTTTTTTTTTTTTTTTTTTTTTTTTTTTTTTT

% ===============================
\subsection{Point-symmetry} 
\label{subsec:PointSymmetry}
% ===============================
Point-symmetry morphological features in CCSNRs are clear predictions of the JJEM. Therefore, the two CCSNRs that I study in this section strongly support the JJEM.
% ===============================
\subsubsection{The Vela SNR} 
\label{subsubsec:Vela}
% ===============================

\begin{figure}[t]
	\centering
%	\hspace*{-2cm} 
	% [trim=left bottom right top, clip]{file}
%	\hspace{1cm}
\includegraphics[trim=0.8cm 10.2cm 3.0cm 2.2cm ,clip, scale=0.50]{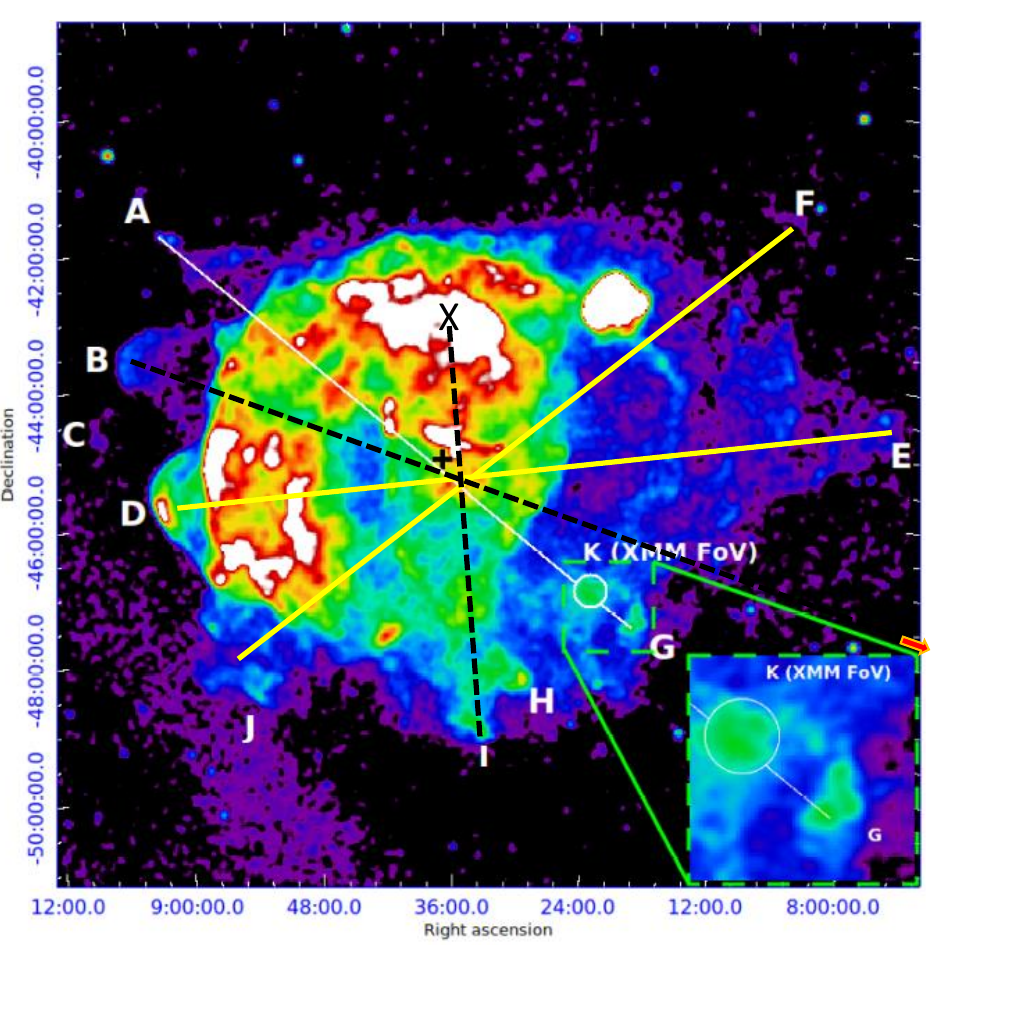} 
\caption{ROSAT X-ray image of SNR Vela \citep{Aschenbachetal1995}, based on figure 1 from \cite{Sapienzaetal2021}. The white \textit{AG-line} and the labelling of the clumps are from their figure (clumps A-F are from \citealt{Aschenbachetal1995}). I added the thick-yellow DE-line and the FJ-line.
I also added two dashed-black lines that connect clumps to my assumed counter jets. 
}
\label{Fig:Vela1}
\end{figure}

The best example of a SNR that contains point-symmetric morphological features is the SNR Vela that I present in Fig. \ref{Fig:Vela1}. This is a ROSAT X-ray image \citep{Aschenbachetal1995} that is based on figure 1 from \cite{Sapienzaetal2021}. The white AG-line is from their figure and was already drawn by \cite{Garciaetal2017}.  
The labelling of the clumps is also from \cite{Sapienzaetal2021}, where clumps A-F were identified by \cite{Aschenbachetal1995}. The high Si abundance of clump A \citep{KatsudaTsunemi2006} and of clumps G and K  \citep{Garciaetal2017} indicates that, as in Cassiopeia A (section \ref{subsec:Ears}), these clumps originate from deep inside the core of the progenitor. \cite{Sapienzaetal2021} convincingly argue that clumps K and G are indeed counter to clump A, and represent jet-like structure from the explosion process.  \cite{KatsudaTsunemi2005} analyze clump D and find it to be overabundance in ONeMg, which suggests that its origin is from near the center of the remnant, as also suggested by \cite{Sankritetal2003}. \cite{GrichenerSoker2017ears} analyze the ears D and E to be the only ears in SNR Vela, and estimate that the combined energy of the jets that inflated ears D and E is only $\approx 1 \%$ of the Vela explosion energy. This is the lowest value among the eight SNRs with ears that they analyze.    
% FFFFFFFFFFFFFFFFFFFFFFFFFFFFFFFFFFFFFFFFFFF  
%FFFFFFFFFFFFFFFFFFFFFFFFFFFFFFFFFFFFFFFFFF

I added to Fig. \ref{Fig:Vela1} the thick-yellow DE-line and the FJ-line, each connecting two previously identified clumps. I here claim that each of the clump pairs AG, DE, and FJ was inflated by one late jet-launching episode during the explosion of Vela. Furthermore, I speculate that the jet that ejected clump B had a counter jet. However, because of the lower density ejecta in the counter-jet-B direction (south-west) this clump moved to larger distances than any other clump, and it is below detection limit. I mark this assumption by a red-orange arrow on the right edge of the figure, and connect it with a dashed-black line to clump B.  In the case of clump I, which I take also to have been formed by a jet, I suggest that the counter-clump(s) is immersed in the large white area in the north. I mark it with a black `X'. { Indeed, \cite{Micelietal2008} identified several shrapnels in that region. \cite{Micelietal2008} find that some of these shrapnels have enhanced Ne and Mg abundances, implying they are ejecta from inner stellar zones. In the JJEM the different compositions of different clumps (shrapnels) suggests that the jets interacted with different layers of the core. The final composition depends on the exact time the jet was launched and how deep it penetrated through inner layers of the core. }  

Overall, in the frame of the JJEM I identify five late jet-launching episodes. There might be more but such that the clumps are projected on the main ejecta of the SNR and therefore are not identified as fast-moving clumps. 
If the energy of these jets are similar to the energy of the jets that inflated ears D and E as \cite{GrichenerSoker2017ears} estimated, then the total energy of the late jets is $\approx 5 \%$ of the explosion energy of Vela. This energy is close to the energy of late jets of CCSNRs that have only one late jet-launching episode (section \ref{subsec:Ears}).   

% ===============================
\subsubsection{SNR 0540-69.3} 
\label{subsubsec:SNR0540693}
% ===============================

\begin{figure*}[ht!]
	%\centering
\includegraphics[trim=0.2cm 13.2cm 0.0cm 1.5cm ,clip, scale=0.75]{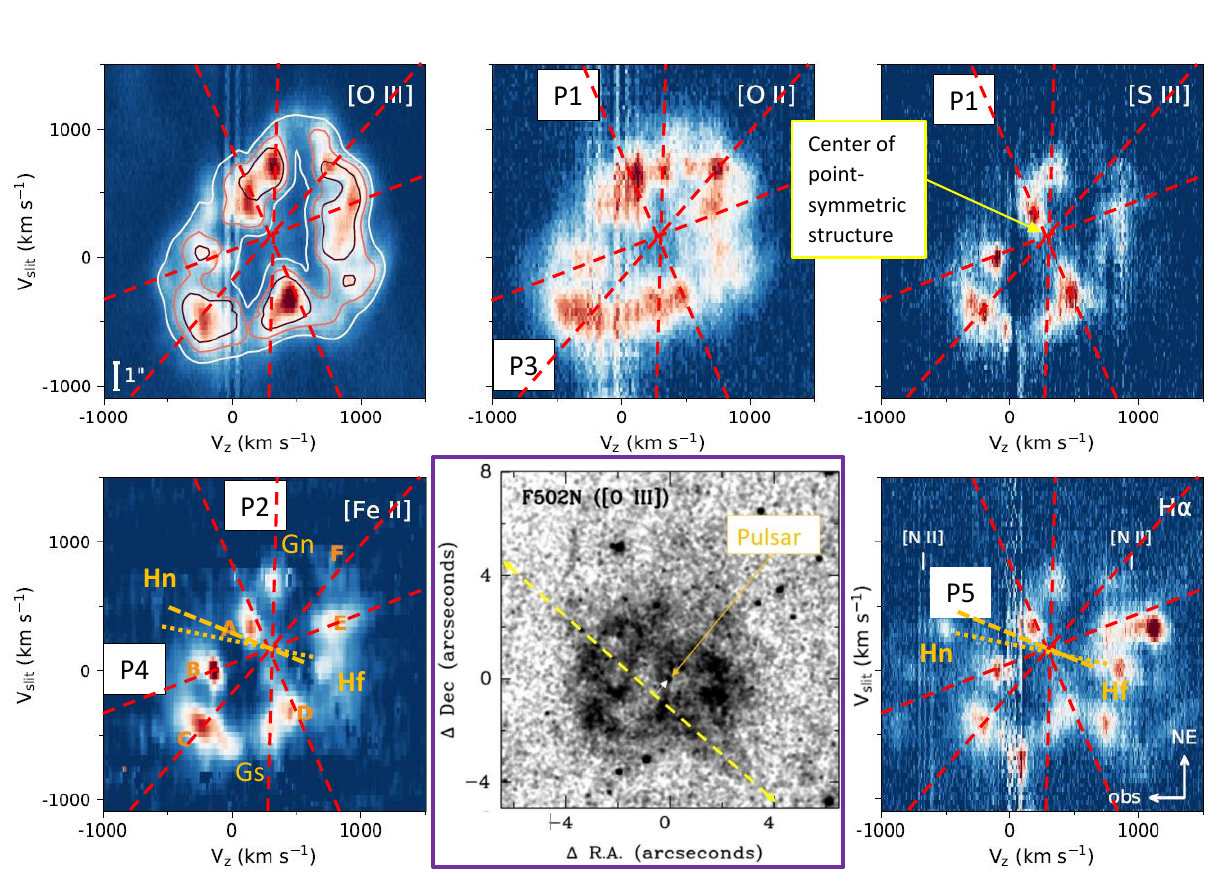} \\
\caption{Five panels of two-dimensional velocity maps of SNR 0540-69.3 based on figure 4 by \cite{Larssonetal2021}. 
The velocities are along a slit that is more or less along the dashed yellow line in the lower-middle panel: $v_{\rm slit}$ is the velocity along the slit (positive to the northeast), while $v_{\rm z}$ is the velocity along the line of sight. The lower-middle panel is an HST image from  \cite{Morseetal2006} to which I added the yellow double-headed arrow. The four dashed-red lines in the five panels that connect opposite clumps are from \cite{Soker2022SNR0540}, where more details can be found.
Clumps A to F are marked by \cite{Larssonetal2021} and clumps Gn and Gs by \cite{Soker2022SNR0540}. I here added the dashed-orange and dotted-orange lines in the two lower velocity maps to indicate another pair, clump Hf and its counter clump Hn. The pulsar is at $v_{\rm slit}=0$ in these panels. }
	\label{fig:SNR0540693}
\end{figure*}

Another SNR with a point-symmetric morphological component is SNR 0540-69.3. I analyzed its point-symmetric morphology \citep{Soker2022SNR0540} as revealed by the detailed observations of \cite{Larssonetal2021}. I present this SNR in  Fig. \ref{fig:SNR0540693}. Five panels are VLT/MUSE velocity maps that \cite{Larssonetal2021} present and which reveal the point-symmetric structure in that plane. This plane is along  the line of sight and through the center of the SNR, more or less along the yellow double-headed arrow in the lower-middle panel of Fig. \ref{fig:SNR0540693}. This panel is an HST observation from \cite{Morseetal2006}.  
% FFFFFFFFFFFFFFFFFFFFFFFFFFFFFFFFFFFFFFF
% FFFFFFFFFFFFFFFFFFFFFFFFFFFFFFFFFFFFFFF

There are four pairs of two opposite clumps in the velocity maps that compose the point-symmetric structure of SNR 0540-69.3.
Unlike the case of SNR Vela where the clumps are at the outskirts of the SNR, in SNR 0540-69.3 the point-symmetric clumps appear in the center of the ejecta (as is evident by their relatively low expansion velocity). I argued in \cite{Soker2022SNR0540} that two to four pairs of jittering jets shaped the inner ejecta in this plane. Here I add another possible pair of clumps as the lines P5 in the lower panels indicates. The clump Hf appears in both the [Fe II] map (lower-left panel) and in the H$\alpha$ map (lower-right panel) at about the same place. The much fainter counter-clump Hn is not exactly at the same place in the two velocity maps. So I draw two lines, the dashed-orange represents the pair in the [Fe II] map and the dotted-orange represents the pair in the H$\alpha$ velocity map. Overall, I here claim for five pairs that form the point-symmetric structure in the velocity maps. 

The lower-middle panel presents a hollowed central region (a faint strip) that connects two ears, the south-west being much longer. The yellow double-headed arrow in the lower-middle panel is along this hollowed region. As the yellow doubled-headed arrow is more or less the direction of the slit that \cite{Larssonetal2021} use for the velocity maps, the pair of ears, which is part of the point-symmetric structure, is in the same plane as the five pairs of clumps that the velocity maps reveal. In \cite{Soker2022SNR0540} I pointed out that the similarity of the point-symmetric structure of SNR 0540-69.3 with some planetary nebulae, e.g., He2-138 (PN~G320.1-09.6; image in  \citealt{SahaiTrauger1998})  and M1-37 (PN~G002.6-03.4; image in \citealt{Sahai2000}), strongly suggests shaping by jets.  

The SNR 0540-69.3 can be classified as point-symmetric with a hollowed-cylinder (barrel-like) structure (more details in \citealt{Soker2022SNR0540}). 
Without the detailed analysis by \cite{Larssonetal2021}, and based only on the HST observations by \cite{Morseetal2006}, this SNR would have been classified as having one-pair of ears. However, while in the SNR Vela the point-symmetric structure is in the outer parts of the ejecta, the velocity maps of SNR 0540-69.3 reveals a point-symmetric structure in the inner parts of the ejecta.  It seems that this inner structure was shaped by the jets that exploded the star. 
Namely, in addition to instabilities in the explosion process (section \ref{sec:intro}) jets also shape the inner ejecta. The jets can play a role in mixing elements in the ejecta of core collapse supernovae.

However, as far as late jets are concerned, I classify SNR 0540-69.3 in the one-pair of ears morphological class.  

% ===============================
\subsection{One pair of ears} 
\label{subsec:Ears}
% =============================== 

CCSNRs that have one pair of ears that dominate their morphology is the largest class. An ear is defined as a protrusion from the main ejecta (nebula) that is fainter than the general nebula, and has a cross section that monotonically decreases from its base on the main nebula to its tip. In most cases the two ears in a pair are not equal in their size and intensity to each other, nor in their distance from the center. The asymmetry is another manifestation of the asymmetrical explosion process of CCSNe that involve instabilities as well as large scale asymmetries. Another prominent manifestation of the asymmetrical explosion is NS natal kick (which I do not study here).

\begin{figure}[t]
	\centering
%	\hspace*{-2cm} 
	% [trim=left bottom right top, clip]{file}
%	\hspace{1cm}
\includegraphics[trim=0.5cm 8.0cm 16.0cm 3.2cm ,clip, scale=0.40]{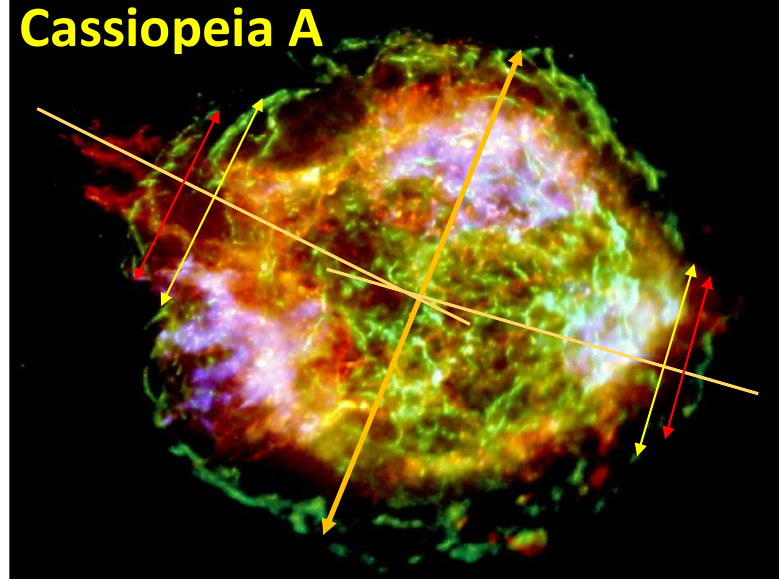}  \\
\includegraphics[trim=2.5cm 16.2cm 1.0cm 2.2cm ,clip, scale=0.41]{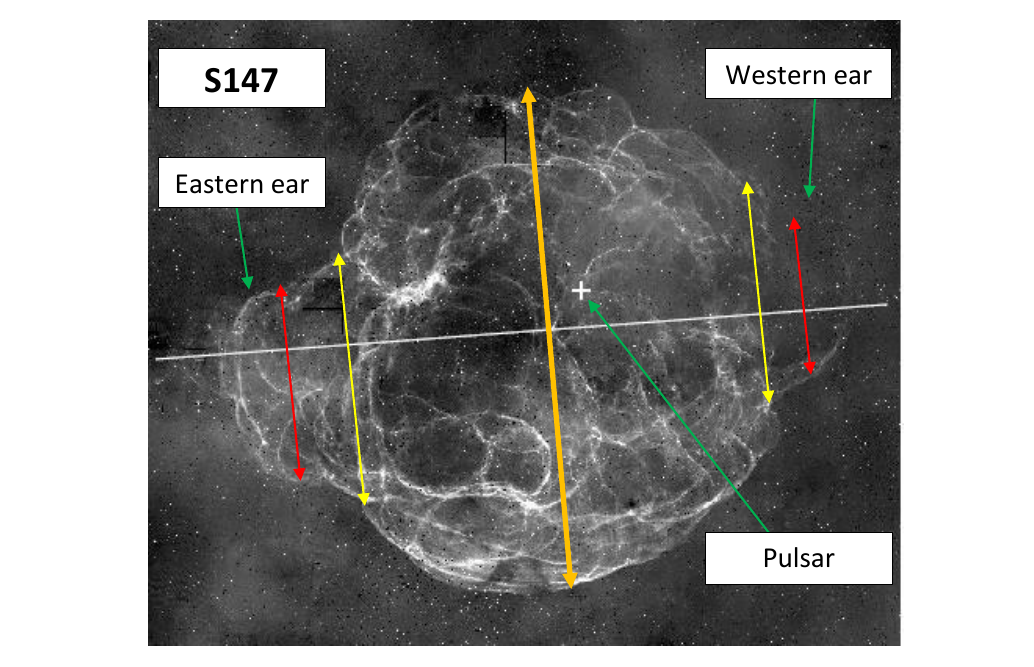} \\
\hspace{0.5cm}
\includegraphics[trim=3.4cm 19.0cm 2.0cm 2.6cm ,clip, scale=0.36]{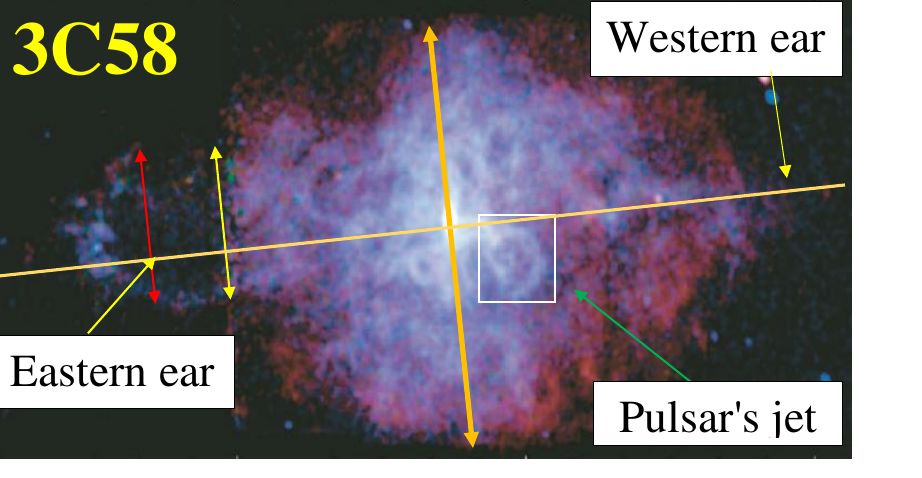}
\caption{Images of three SNRs where one pair of ears dominate the outer morphology, and where at least one ear is large and prominent. Upper three images: The identification of the ears and the double-headed arrow marks of the base of an ear at the main ejecta and of the center of an ear are from \cite{GrichenerSoker2017ears}.  
The sources of the images are as follows. 
\textit{Cassiopeia A:} An X-ray image taken from the Chandra gallery (based on \citealt{Hwangetal2004}).
% Red, blue and green represent Si He$\alpha$ (1.78-2.0 keV), Fe K (6.52-6.95 keV), and  4.2-6.4 keV continuum, respectively. 
%
 \textit{S147:} An H$\alpha$ image from \cite{Gvaramadze2006} who reproduced an image from \cite{Drewetal2005}.
 \textit{3C58:}  ACIS/Chandra image from the Chandra Gallery based on \cite{Slaneetal2004}; colors represent energy bands. 
% $0.5 - 1.0 \keV$ (red), $1.0-1.5 \keV$ (green), and $1.5-10 \keV$ (blue). 
}
\label{Fig:CassiopeiaA}
\end{figure}

\begin{figure}[t]
	\centering
%	\hspace*{-2cm} 
	% [trim=left bottom right top, clip]{file}
%	\hspace{1cm}
\includegraphics[trim=2.4cm 14.5cm 2.0cm 2.0cm ,clip, scale=0.55]{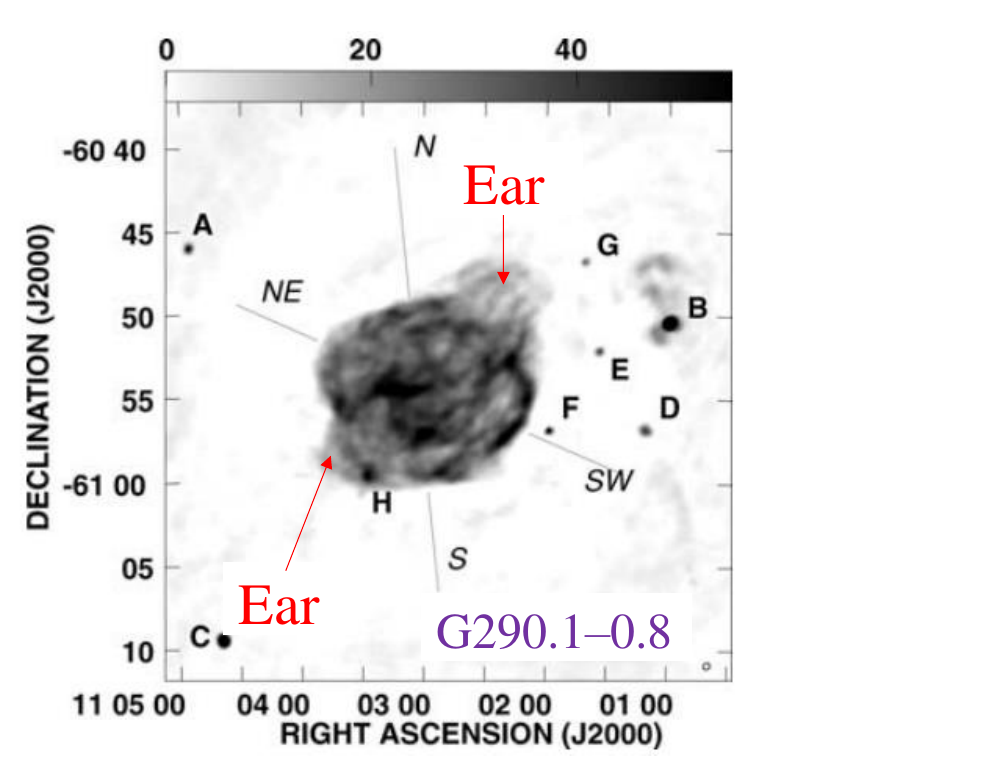}
\caption{Radio continuum image at 1384 MHz of SNR G290.1$-$0.8 that morphologically belongs to SNRs in Fig. \ref{Fig:CassiopeiaA}. From \cite{Reynosoetal2006} to which I added the identification of ears. 
% $0.5 - 1.0 \keV$ (red), $1.0-1.5 \keV$ (green), and $1.5-10 \keV$ (blue). 
}
\label{Fig:G290}
\end{figure}

\begin{figure}[t]
	\centering
%	\hspace*{-2cm} 
	% [trim=left bottom right top, clip]{file}
%	\hspace{1cm}
\includegraphics[trim=0.7cm 10.0cm 4.0cm 6.2cm ,clip, scale=0.345]{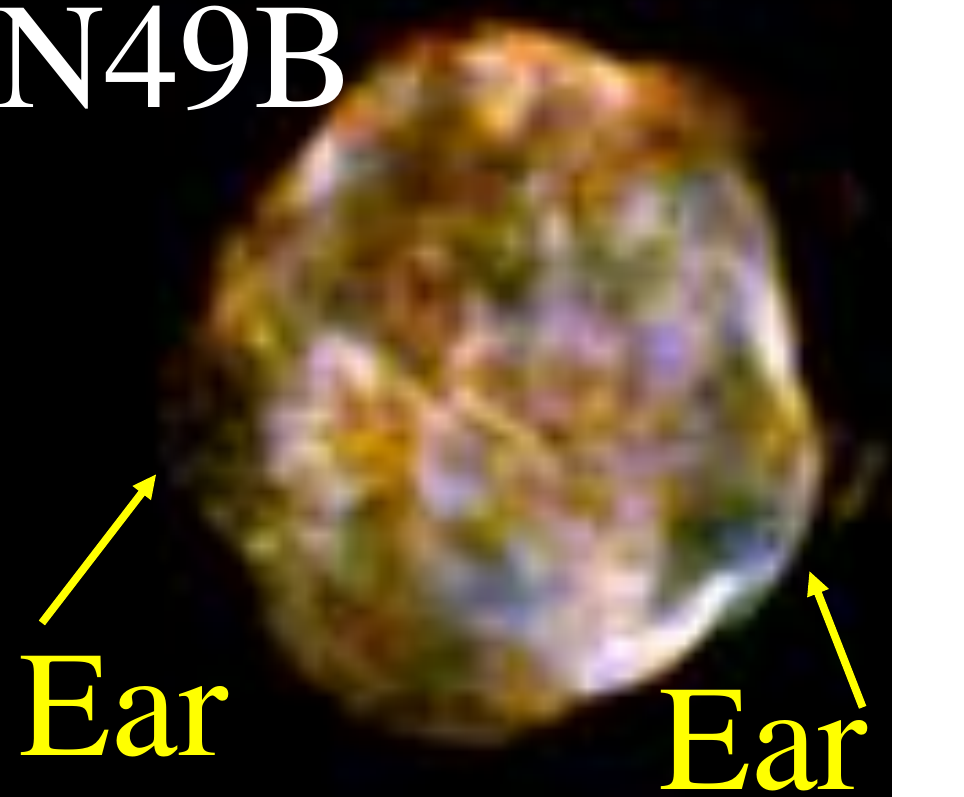} \\
\includegraphics[trim=3.55cm 10.0cm 4.0cm 4.0cm ,clip, scale=0.535]{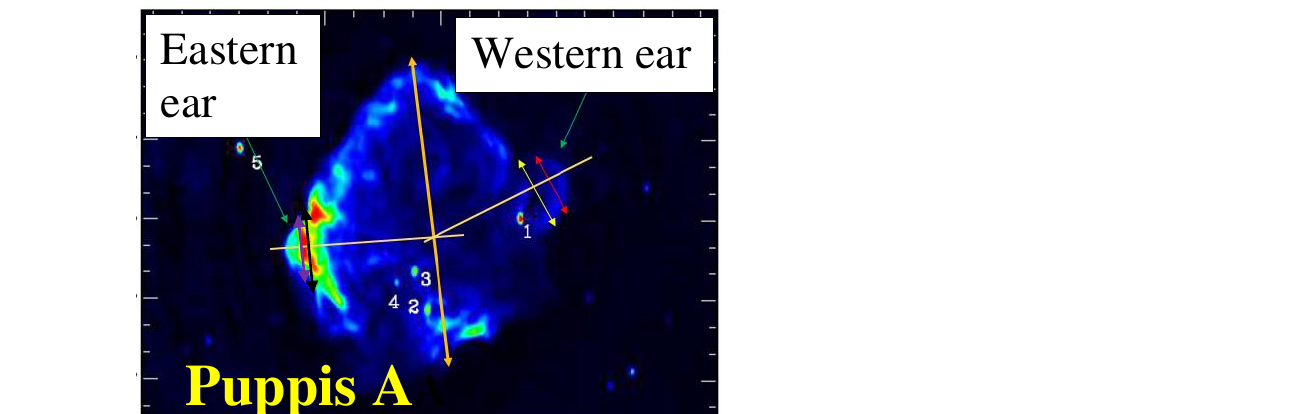}
\includegraphics[trim=3.85cm 14.0cm 2.0cm 4.5cm ,clip, scale=0.376]{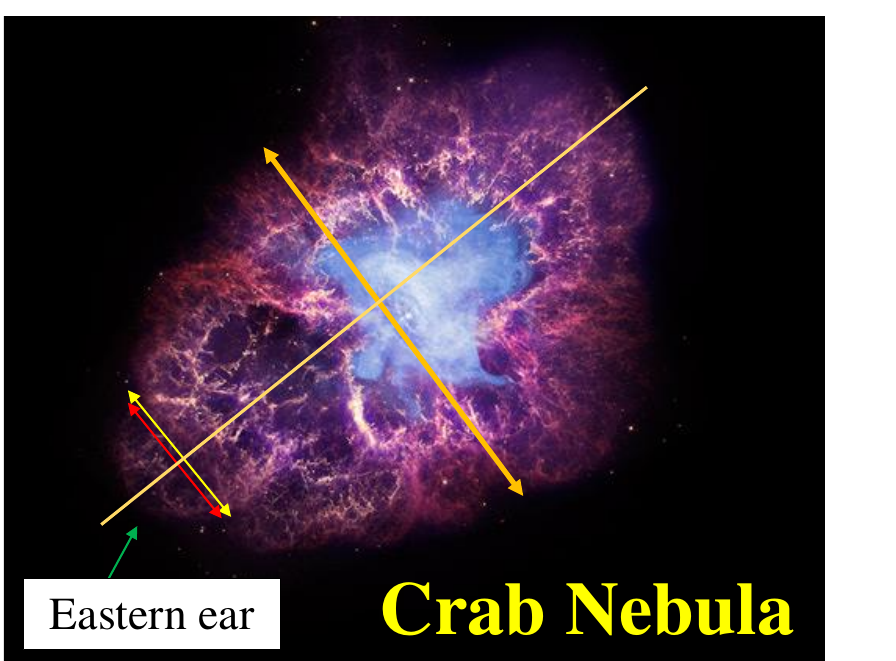}
\caption{Images of three SNRs with one pair of ears that do not protrude much from the main ejecta. Sources of marks in the two lower panels are from \cite{GrichenerSoker2017ears}.   
The sources of the images are as follows. 
 \textit{N49B:} An X-ray image from the Chandra gallery based on \cite{Parketal2003}.
 \textit{Puppis A:} The radio continuum emission at 1.4 GHz; published by \cite{ReynosoWalsh2015} and reproduced by \cite{Reynosoetal2017}.
 \textit{Crab Nebula:} A composite image of X-ray (Blue; \citealt{Sewardetal2006}), Optical (Red-Yellow; \citealt{Hester2008}) and Infrared (Purple; NASA/JPL-Caltech/Univ). 
}
\label{Fig:N49B}
\end{figure}

\cite{GrichenerSoker2017ears} and \cite{Bearetal2017} study many of these CCSNRs and estimated the extra energy of the jets that inflated the pair of bubbles. These studies find that the extra energy varies between different CCSNRs, from being $\simeq 1 \%$ to $\simeq 30 \%$ of the total explosion energy. I here examine only the morphology. 
In Figs. \ref{Fig:CassiopeiaA} - \ref{Fig:N49B} I present seven images, mostly from \cite{GrichenerSoker2017ears} who marked with double-headed arrows the base and middle of the ears. 
% FFFFFFFFFFFFFFFFFFFFFFFFFFFFFFFFFFFFFFFFFFF  
% FFFFFFFFFFFFFFFFFFFFFFFFFFFFFFFFFFFFFFFFFF
% FFFFFFFFFFFFFFFFFFFFFFFFFFFFFFFFFFFFFFFFFFF  
% FFFFFFFFFFFFFFFFFFFFFFFFFFFFFFFFFFFFFFFFFF
% FFFFFFFFFFFFFFFFFFFFFFFFFFFFFFFFFFFFFFFFFFF  
%FFFFFFFFFFFFFFFFFFFFFFFFFFFFFFFFFFFFFFFFFF

One of the best example of the one-pair class is S147 that I also present in Fig. \ref{Fig:CassiopeiaA} (for a recent study of this SNR see, e.g., \citealt{Renetal2018}). The two other SNRs in Fig. \ref{Fig:CassiopeiaA} and the one in Fig. \ref{Fig:G290} have one ear much larger than the other. Fig. \ref{Fig:N49B} present three SNRs with ears that do not protrude much from the main ejecta (nebula).  

% ==========================================
\subsection{S-shaped morphology} 
\label{subsec:Sshape}
% ==========================================

\begin{figure}[t]
	\centering
%	\hspace*{-2cm} 
	% [trim=left bottom right top, clip]{file}
%	\hspace{1cm}
 \includegraphics[trim=3.7cm 14.0cm 1.0cm 2.4cm ,clip, scale=0.52]{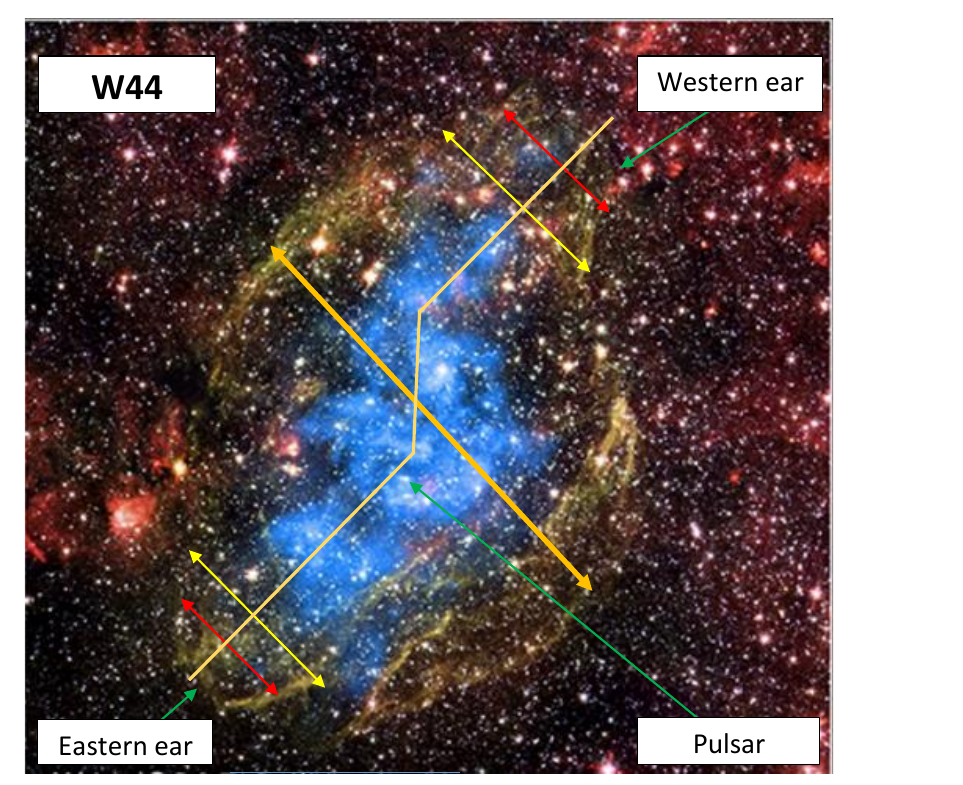} 
\caption{A composite image of SNR W44 taken from the Chandra gallery with marks from \cite{GrichenerSoker2017ears}. The cyan color represents X-ray (based on \citealt{Sheltonetal2004}). The red, blue and green represent infrared emission (based on NASA/JPL-Caltech). This SNR has a prominent S-shaped morphology.   
}
\label{Fig:W44}
\end{figure}

This class includes only the SNR W44 that I present in Fig. \ref{Fig:W44} taken from the Chandra gallery with lines from \cite{GrichenerSoker2017ears}. The S-shaped morphology is most likely due to precession of the jets around a fixed axis.
The two ears are not symmetric nor with respect to the Pulsar and nor with respect to the main shell. 
% FFFFFFFFFFFFFFFFFFFFFFFFFFFFFFFFFFFFFFFFFFF  
%FFFFFFFFFFFFFFFFFFFFFFFFFFFFFFFFFFFFFFFFFF

The morphology of W44 is of one pair of ears that is arranged in an S-shape. It can be as well belong also to the one-pair class. However, the very likely cause of an S-shape is jet-precession. Namely, it was the accretion disk that launched the last jets that performed precession while launching the jets. This suggests, in the fame of the JJEM, a non-negligible pre-collapse core rotation as I discuss in section \ref{sec:CoreRotation}. 

% ==========================================
\subsection{Barrel-shaped SNRs.} 
\label{subsec:Barrel}
% ==========================================

\begin{figure}[t]
	\centering
%	\hspace*{-2cm} 
	% [trim=left bottom right top, clip]{file}
%	\hspace{1cm}
\includegraphics[trim=0.2cm 16.7cm 2.0cm 2.2cm ,clip, scale=0.49]{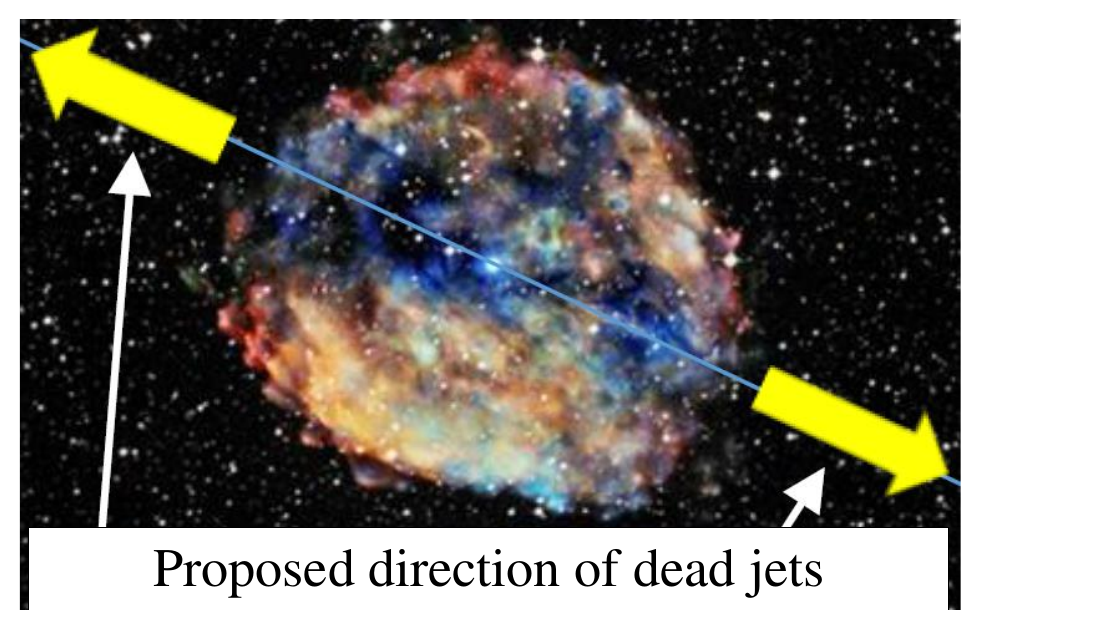}  
\caption{An X-ray image of RCW 103 in three energy bands (low=red, medium=green, highest=blue) combined with an optical image from the Digitized Sky Survey (image taken from the Chandra website based on \citealt{Reaetal2016}). 
The yellow arrows mark the original directions of the already dead jets as \cite{Bearetal2017} proposed. } 
\label{Fig:RCW103}
\end{figure}

A barrel-shaped morphology refers to a general axisymmetrical structure with a central region along the symmetry axis that is much fainter than the sides. The two ends on the symmetry axis are trimmed. Its hollowed structure appears in observations as two opposite bright arcs with a faint (hollowed) region between them. The best example of a barrel-shaped SNR is RCW 103 that I present in Fig. \ref{Fig:RCW103}. I take this X-ray image \citep{Reaetal2016} from \cite{Bearetal2017} who proposed the shaping of RCW 103 by two jets at the final phase of the explosion. They based the jet-shaping model on the morphological similarities of RCW 103 with several barrel-shaped planetary nebulae that are observed to be shaped by jets. The unequal structure of the two arcs, which are the projection of the barrel-structure on the plane of the sky, can result from a density gradient in the interstellar medium (e.g.,  \citealt{Luetal2021}) or from asymmetrical explosion. 
% FFFFFFFFFFFFFFFFFFFFFFFFFFFFFFFFFFFFFFFFFFF  
%FFFFFFFFFFFFFFFFFFFFFFFFFFFFFFFFFFFFFFFFFF

\begin{figure} %[H]
\centering
%\hskip -6.00 cm
\includegraphics[trim=9.4cm 3.5cm 0.0cm 3.0cm ,clip, scale=0.58]{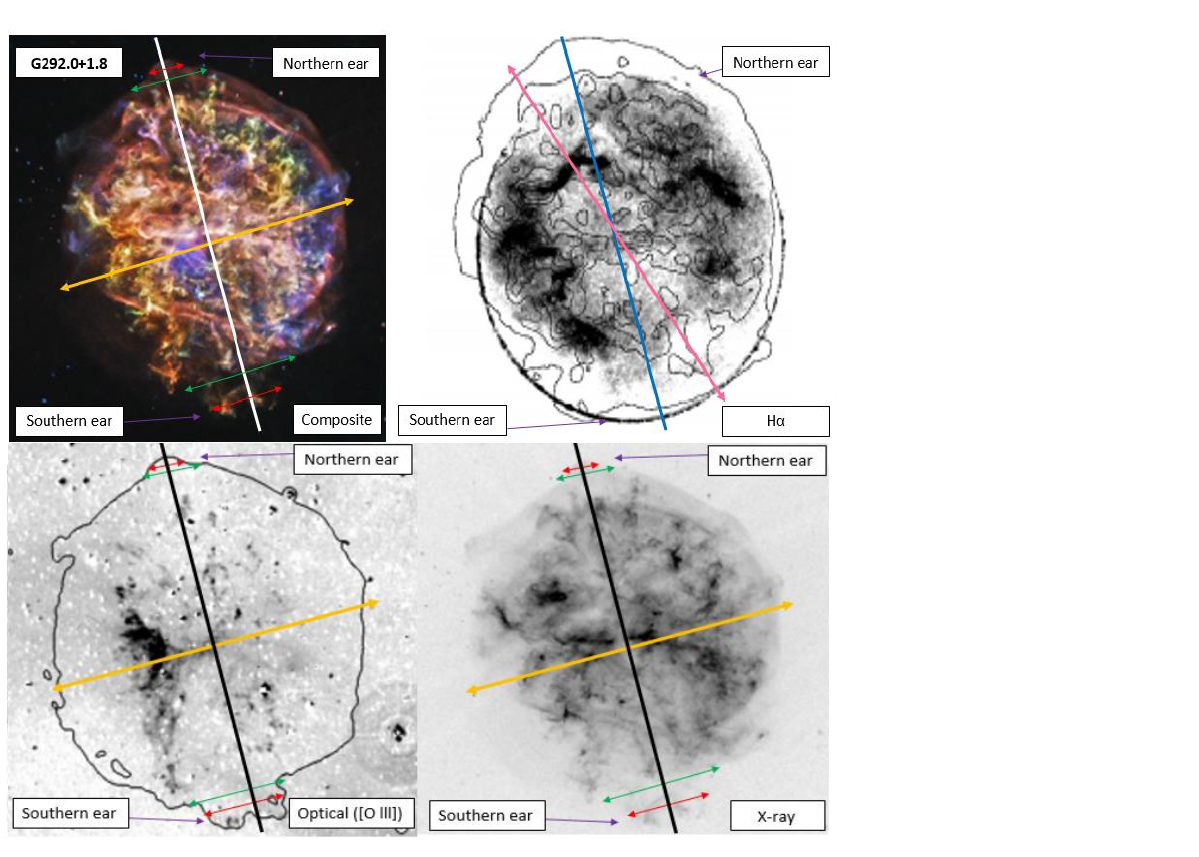}  
%\vskip -2.00 cm
\caption{Images of the CCSNR G292.0+1.8 in various wavelengths with marks from \cite{Bearetal2017}. In each image there is a line that connect the two opposite ears that \cite{Bearetal2017} define and analyze. On the H$\alpha$ image they also define the symmetry axis of the barrel-shaped morphology by the double-headed pink line. 
    \textit{Upper left panel:} A composite X-ray image \citep{Parketal2007} from the Chandra gallery where different lines represent different energy bans (for another X-ray image see \citealt{Yangetal2014}).  
    %The red, orange, green and blue colors represent O~Ly$\alpha$ combined with Ne~He$\alpha$ ($0.58 - 0.71 \keV$ and $0.88 - 0.95 \keV$), Ne~Ly$\alpha$ ($0.98 - 1.10 \keV$), Mg~He$\alpha$ ($1.28-1.43 \keV$), and Si~He$\alpha$ combined with S~He$\alpha$ ($1.81 - 2.05$ and $2.40 - 2.62 \keV$), respectively. White represents the optical band.   
    \textit{Upper right panel:} Zero velocity H$\alpha$ image taken from \cite{Ghavamian2005}, which clearly reveals the barrel-shaped morphology.  
    \textit{Lower left panel:} An optical ([O III]) image taken from \cite{WinklerLong2006} and reproduced by \cite{Ghavamian2012}. 
    Lower right panel: A Chandra $0.3-8.0 \keV$ X-ray image based on  \cite{Parketal2007} and reproduced by \cite{Ghavamian2012}. } 
\label{Fig:G292.0+1.8}
\end{figure}

The case of SNR G292.0+1.8 is subtle as it shows both a barrel-shaped morphology and two opposite ears. In Fig. \ref {Fig:G292.0+1.8} I present an image from \cite{Bearetal2017} where more details can be found. The visible images of H$\alpha$ (upper-right panel) and [O III] (lower-left panel) show the barrel-shaped morphology. \cite{Bearetal2017} indicate the symmetry axis of the barrel-shaped morphology by the double-headed pink line in the H$\alpha$ image. The X-ray images, on the other hand, present two very small opposite ears that \cite{Bearetal2017} mark and analyze. Because the two opposite arcs in the H$\alpha$ image present a much prominent barrel-shaped morphology than the two small ears, I classified it as barrel-shaped SNR. 
%FFFFFFFFFFFFFFFFFFFFFFFF
%FFFFFFFFFFFFFFFFFFFFFFFF

\begin{figure}[]
\centering
\includegraphics[trim=6.4cm 14.5cm 0.0cm 2.0cm ,clip, scale=0.58]{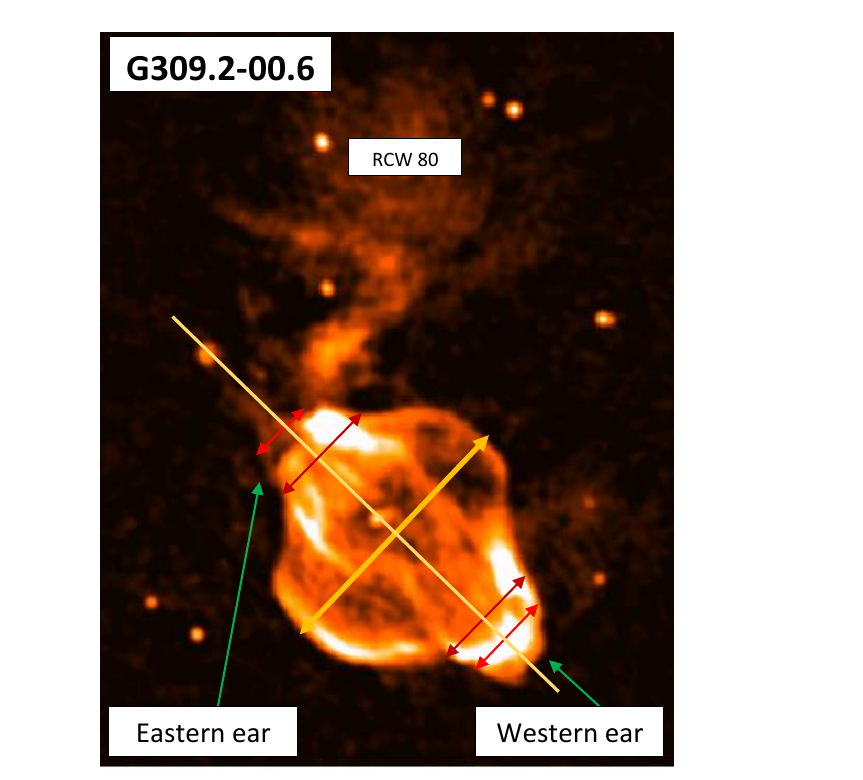}
\caption{A radio image of SNR G309.2-00.6 from the site of the School of Physics, The university of Sydney (posted as production from \citealt{Gaensleretal1998}).  
Marks are from \cite{GrichenerSoker2017ears}. 
In the background is the emission nebula RCW 80. 
}
 \label{Fig:G3092006}
\end{figure}

SNR G309.2-00.6 that I present in Fig. \ref{Fig:G3092006} with marks from \cite{GrichenerSoker2017ears} also presents a complicated case. It has two prominent ears as marked on the figure. However, in addition there is a hollowed zone along the symmetry axis (yellow line). The sides of the symmetry axis present two opposite arcs on the outskirts of the ejecta which complicate the morphology. 
I classify it as barrelled-shape SNR. No NS was found in this SNR, but its morphology and location in the Galaxy strongly suggest a CCSN origin \citep{Gaensleretal1998}. If, as I argue in section \ref{sec:CoreRotation}, the progenitor core was rapidly rotating it might have collapsed to a BH (see also section \ref{subsec:Elongated}).  
%FFFFFFFFFFFFFFFFFFFFFFFF
%FFFFFFFFFFFFFFFFFFFFFFFF

\cite{YuFang2018} showed by hydrodynamical simulations that jets with a total energy of $\simeq 10-15 \%$ of the explosion energy can shape the morphology type of SNR G309.2-00.6. 

{ The CCSNR G156.2+5.7 presents an interesting morphology. Its radio morphology with the polarization structure (magnetic fields) has a clear barrel-shaped morphology as the thorough observation and analysis by \cite{XuJWetal2007} reveal. However, its H$\alpha$ (e.g., \citealt{GerardyFesen2007}) and X-ray (e.g., \citealt{PannutiAllen2004}) images do not possess a barrel-shaped morphology (see comparison of images by \citealt{XuJWetal2007}).  It is a relatively old CCSNR, a few tens of thousands years \citep{Katsudaetal2016}. Therefore, most likely the interaction with the interstellar medium played a major role in shaping its present morphology. For these reasons I do not classify it in this study. }

% ==========================================
\subsection{Elongated SNRs} 
\label{subsec:Elongated}
% ==========================================

\begin{figure}[]
\centering
\includegraphics[trim=0.0cm 11.5cm 0.0cm 1.4cm ,clip, scale=0.39]{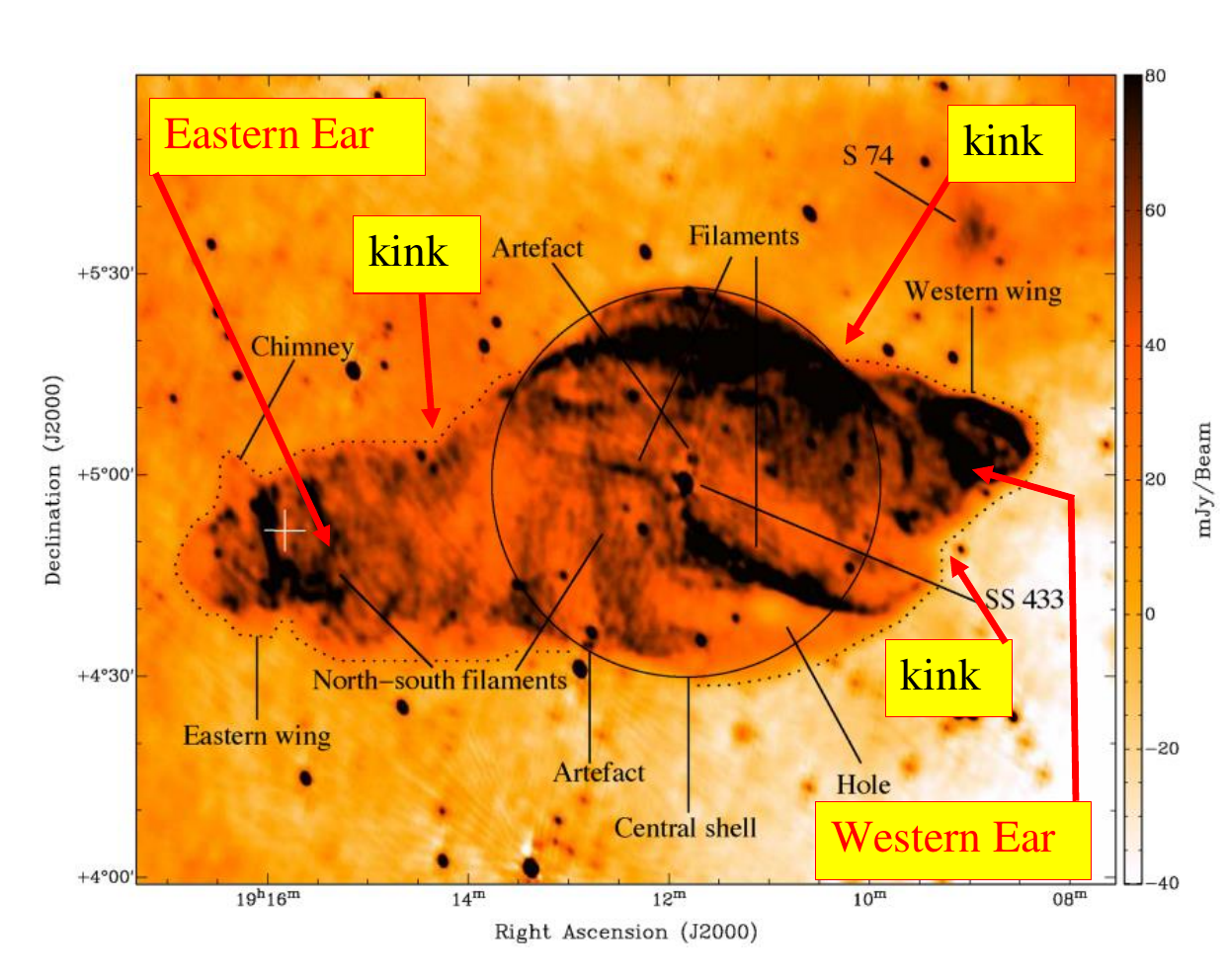} \\
\includegraphics[trim=0.0cm 14.5cm 0.0cm 1.4cm ,clip, scale=0.39]{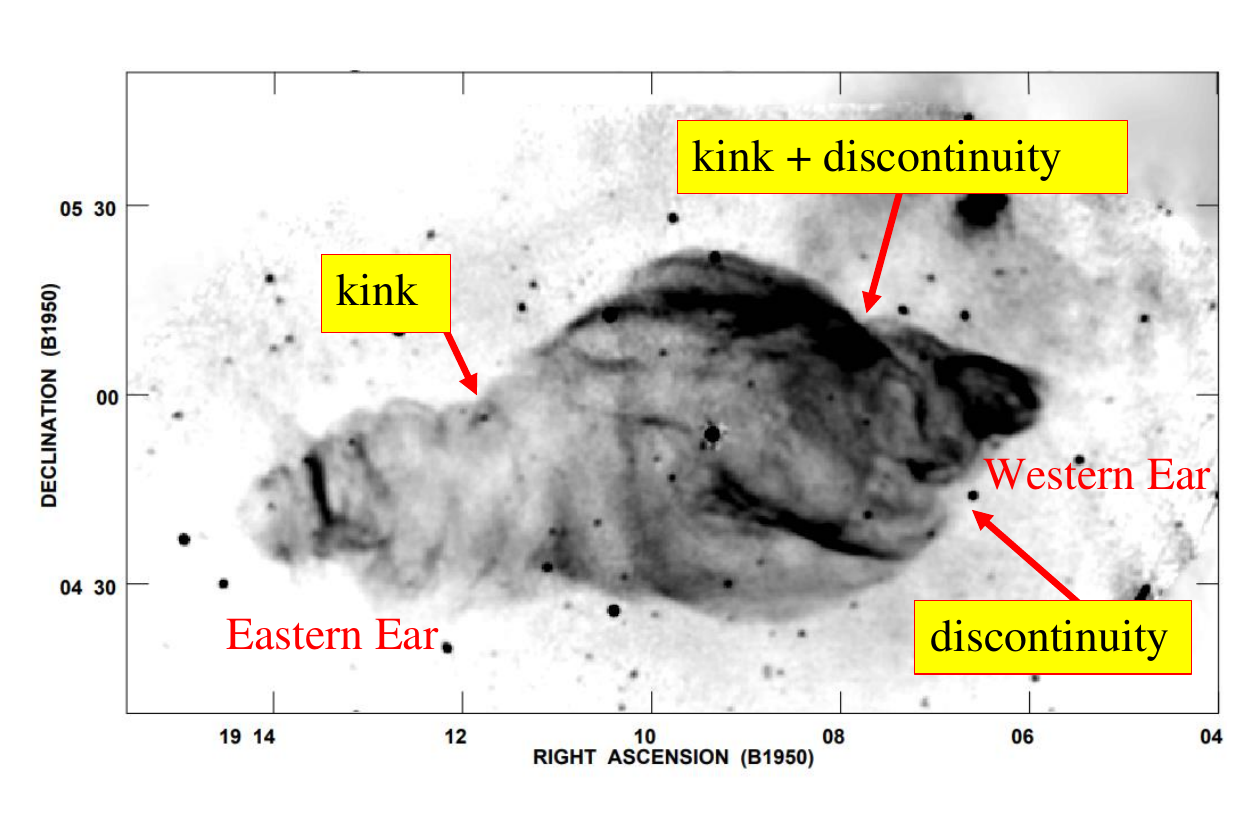}
\caption{Upper panel: A LOFAR 140-MHz high-band continuum map of SNR W50 from \cite{Brodericketal2018}.  
Colour scale runs from $-40~$mJy/beam to $80~$mJy/beam. Most marks are on the original image from \cite{Brodericketal2018} . I added the marks of `kink' for the projected boundaries between the nebula and the ears.    
Lower panel: The SNR W50 in radio continuum at 1465 MHz as observed with the VLA (from \citealt{Dubneretal1998}). I added the marks of `kink' and `discontinuity' in the projected boundaries between the main nebula and the ears. 
}
 \label{Fig:W50}
\end{figure}

The fifth class is of an elongated morphology that only SNR W50 belongs to. However, there are large uncertainties because of the shaping by the jets that its central binary system SS 433 launches and that are not related to the exploding jets. Specifically, the BH component of the binary systems launches these jets. In Fig, \ref{Fig:W50} I present its LOFAR image that I take from \cite{Brodericketal2018} and its VLA radio continuum map from \cite{Dubneretal1998}. I added to these two figures only what I identify as the boundaries between each ear and the main nebula by `kink' and `discontinuity'. Note that in two places the LOFAR image reveals a kink between the surface of the main nebula an the surface of the western ear, while the VLA image also shows a discontinuity between the two surfaces.  
These images show that although the two ears of W50 are connected to the main nebula with small variations between the main nebula and the ears, there is still a clear boundary between the nebula and the ears. 
%FFFFFFFFFFFFFFFFFFFFFFFF
%FFFFFFFFFFFFFFFFFFFFFFFF

\cite{Ohmuraetal2021} argue that the continuous jets from SS 433 formed the entire W50 nebula. The shocked material of the jets and of the interstellar medium (ISM) into which the jets propagate, i.e., the cocoons, formed the main nebula (the central part). The fronts of the jets form the ears. In their scenario SS 433 has been launching the jets for the last $\simeq 10^5 \yr$. The problem I find with their model is that their morphology do not reproduce the clear boundaries between the main nebula and the two ears because the jets produce both the main nebula and the ears. Specifically, they do not reproduce the `kinks' and the `discontinuities' that I mark on Fig. \ref{Fig:W50}.   
\cite{Goodalletal2011}, on the other hand, do consider W50 main nebula to be a SNR. They conduct hydrodynamical simulations where they launch jets that the BH in SS 433 launches into a spherical supernova remnant. They obtain clear ears with clear boundaries from the main nebula.
The problem I find with the images that \cite{Goodalletal2011} obtain is that the ears largely differ from the main nebula, much more than observed. 

The hydrodynamical simulation results of \cite{Ohmuraetal2021} that the ears are basically part of the main nebula, more than observed in W50, and of \cite{Goodalletal2011}, that the ears differ from the main nebula to much larger degree than observed in W50, bring me to suggest an intermediate scenario.  I take these results to imply that the ears were created during the jet-driven explosion process of W50 and were further shaped by the later jets that the system SS 433 has been launching.    

In section \ref{sec:CoreRotation} I discuss the theoretical motivation to introduce the elongated class of SNRs.

% ==========================================
\section{The possible role of core rotation} 
\label{sec:CoreRotation}
% ==========================================

In the JJEM there are two sources of the angular momentum of the mass that the newly born NS accretes.
This is true also in cases where the NS collapses to a BH.  
The first angular momentum source is the pre-collapse stochastic convection motion in the collapsing core that introduces angular momentum fluctuations with varying magnitudes and directions. The angular momentum fluctuations due to the core convective motion are amplified by instabilities in the zone between the newly born NS and the stalled shock at $\simeq 100 \km$ from the NS (section \ref{sec:intro}). The other angular momentum source is the pre-collapse core rotation. It introduces an angular momentum component with a fixed direction. Its magnitude slowly increases with time as material from outer layers in the core are accreted. 

In \cite{Soker2023gap} I built a toy model to study the effects of these two angular momentum components on the direction of the jets. I used that toy model to offer an explanation to the $\simeq 2.5 - 5 M_\odot$ mass gap between NSs and BHs in the frame of the JJEM. I assumed in that toy model that all specific angular momentum fluctuations of the random angular momentum component, after amplification by post-shock instabilities, have the same magnitude of $j_{\rm f}$ and have stochastically direction variations. I took the typical range of values to be $j_{\rm f} \simeq 2 \times 10^{16} \cm^2 \s^{-1} - 5 \times 10^{16} \cm^2 \s^{-1}$. The pre-collapse core rotation introduces a fixed-direction specific angular momentum component of magnitude $j_{\rm p}$.  
I found with the above toy model that when the core is slowly rotating, $j_{\rm p} \la 0.5  j_{\rm f}$, the jets are launched in all directions. According to the JJEM in this case the jet feedback mechanism is efficient and the jets explode the core early-on, leaving a NS remnant 
(e.g., \citealt{ShishkinSoker2022}). When the pre-collapse core is rapidly rotating with $j_{\rm p} \gtrsim j_{\rm f}$ the NS does not launch jets in the equatorial plane of the pre-collapse rotating core (the plane perpendicular to $\overrightarrow{j_{\rm p}}$) and its vicinity. The jets do not expel mass efficiently from the equatorial plane and accretion proceeds to form a BH. 
The BH might launch relativistic jets. Such jets might lead to new processes in the supernova that do not occur when a NS is formed, e.g., neutrino emission as in choked gamma-ray bursts (as calculated by, e.g., \citealt{SahuZhang2010, Heetal2018, Fasanoetal2021, Guettaetal2023}). 

The case with $j_{\rm p} \gtrsim j_{\rm f}$, therefore, both maintains a more or less fixed-axis direction of the jets and leaves a BH remnant. The fixed-axis jets form an elongated structure. This is the theoretical motivation behind the morphological class of elongated nebulae (section \ref{subsec:Elongated}), and in classifying W50, which has a BH in its central binary system, in this class. As discussed in section \ref{subsec:Elongated}, in the case of W50 the jets that the binary system SS 433 has been launching further shaped the ears. 

In \cite{Soker2023gap} I studied only the mass gap between NSs and BHs. I did not study the different cases of $j_{\rm p} \lesssim j_{\rm f}$ that leave a NS remnant. I now do that in relation to the first four classes in Table \ref{table:SNRs}. 

When the pre-collapse core rotation plays no role, namely $j_{\rm p} \ll 0.1 j_{\rm f}$, the jets fully jitter at all jet-launching phases. Here I crudely estimate this range as $j_{\rm p} \lesssim 0.01 j_{\rm f}$. The exact value should be determined in the future by highly-demanding three-dimensional hydrodynamical simulations.    
In these cases the end period of the mass accretion process onto the newly born NS can be composed of several short, each lasting $\approx 0.01 \s$,  jet-launching episodes that leaves a point-symmetric structure in the outer regions of the ejecta. This is the case of SNR Vela (Fig. \ref{Fig:Vela1}; section \ref{subsec:PointSymmetry}). 

When the pre-collapse core rotation is somewhat larger it might act to increase the probability of the jets' axis to be close to the angular momentum axis of the pre-collapsing core, i.e., along $\overrightarrow{j_{\rm p}}$. This might cause the last jet-launching episode to be somewhat longer and to form one dominant pair of opposite ears. The last jet-launching episode lasts for a relatively long time because of the following consideration. An accretion disk without fresh supply of material lives for about the viscous timescale of the disk. This can be tens to hundreds times the orbital period of the material. During the explosion process in the JJEM, newly accreted matter has different angular momentum direction than the existing disk and it can destroy the disk. Namely, the freshly accreted material terminates the jets and starts a new jet-launching episode. The last accretion episode in the JJEM has no fresh supply of material. The accretion disk can live for the viscous time scale. For a NS of mass $M_{\rm NS}=1.4 M_\odot$ and an accretion disk at $r=30 \km$ the orbital period of the material is $0.0024 \s$. The viscous time scale might be $\approx 0.1-1 \s$.  
This is a relatively long time (as a regular jet-launching episode lasts for $\approx 0.01-0.1 \s$) during which the outer core expands and the final material of these last jets shape the ears in the expanding core and envelope. 
I therefore suggest that for the range of $j_{\rm p} \approx 0.01 j_{\rm f} - 0.1 j_{\rm f}$ (admittedly this range is a crude estimate), the last jets form a prominent pair of ears, e.g., the one-pair morphology. The final accretion disk might precess due to perturbations by accreted parcels of material, leading to an S-shaped morphology.  

When the pre-collapse core angular momentum is larger, but not as to form a BH, the last jet-launching episode might be longer and more powerful. The jets can clear the central zone around the core angular momentum axis and form a barrel-like morphology. I crudely take this range to be $j_{\rm p} \approx 0.1 j_{\rm f} - 0.3 j_{\rm f}$.

These ranges are crude estimates within the frame of the toy model. The situation is more complicated as the specific angular momentum fluctuations do not have a constant magnitude as the toy model assumes. 

{ I note that the final angular momentum of the NS does not relate monotonically to the pre-collapse core rotation. The reason is that in the JJEM the jets of each jet-launching episode carry most of the angular momentum of the accretion disk that launches the jets. In a case of a rapid pre-collapse rotation there might be one long-lived jet-launching episode with a fixed jets' axis. However, in that case the magnetic fields in the NS and in the accretion disk might very efficiently slow down the NS by coupling the NS to outer disk radii where angular velocity is much slower. Further more, after accretion ceases rapidly rotating NSs substantially slow-down by blowing winds (e.g., \citealt{Prasannaetal2022}) in the propeller mechanism (e.g., \citealt{Ottetal2006}). Therefore, in most, but not in all, cases the JJEM mechanism expect for a spin-period of tens of milliseconds shortly after explosion (e.g., \citealt{GofmanSoker2020}). }

The main point to take from this section is that in the frame of the JJEM the pre-collapse core rotation, or more specifically the ratio $j_{\rm p}/j_{\rm f}$, is the main parameter that determine the outer large-scale morphology of CCSNRs. Other factors are the non-linear instabilities that occur during the explosion, the possible presence of a binary companion, a circumstellar material into which the ejecta expand (e.g., \citealt{Velazquezetal2023}), the energy of the explosion and the ejecta mass, and the interstellar medium (in particular with a strong magnetic field, e.g., \citealt{Wuetal2019, Velazquezetal2023}). 

% ==========================================
\section{Summary} 
\label{sec:Summary}
% ==========================================

I classified 14 CCSNRs into five classes according to morphological features that late jets in the explosion process might form (Table \ref{table:SNRs}). According to the JJEM, after the early jets explode the core the late jets that interact with the already expanding star might leave imprints on the ejecta, outer and inner regions (e.g., \citealt{GrichenerSoker2017ears, Bearetal2017}). 

Late jittering jets where more than one pair of jets leave imprints on the ejecta shape a point-symmetric morphology (Fig. \ref{Fig:Vela1}). I attribute this type of shaping to cases where the pre-collapse core rotation is extremely slow. Namely, the specific angular momentum in the relevant layer of the core due to its rotation is much smaller than the typical magnitude of the specific angular momentum fluctuations due to pre-collapse core convection. Based on earlier results \citep{Soker2023gap} I crudely estimate this range to be $j_{\rm p} \lesssim 0.01 j_{\rm f}$, as I list in the second row of Table \ref{table:SNRs} and discuss in section \ref{sec:CoreRotation}. 

More rapidly rotating cores might force the last pair of jets to be long-lived and shape one pair of jet-inflated ears that dominate the morphology (Figs. \ref{fig:SNR0540693}, \ref{Fig:CassiopeiaA}, \ref{Fig:G290}, and \ref{Fig:N49B}). The accretion disk might precess, therefore leading to an S-shaped morphology (Fig. \ref{Fig:W44}). I crudely estimate that these cases occur when $j_{\rm p} \approx 0.01 j_{\rm f} - 0.1 j_{\rm f}$. 
Even more rapidly rotating pre-collapse cores, which I crudely estimate to have  $j_{\rm p} \approx 0.1 j_{\rm f} - 0.3 j_{\rm f}$, might clear the region along the axis of the pre-collapse core rotation and form a barrel-shaped morphology (Figs. \ref{Fig:RCW103}, \ref{Fig:G292.0+1.8} and \ref{Fig:G3092006}). 

The most uncertain class of this study is the elongated morphology which includes only SNR W50 that has a binary system that launches jets (Fig. \ref{Fig:W50}). I argued in section \ref{subsec:Elongated} that both the exploding jets and the jets that the BH in the binary system launches have shaped the ears of W50. This class occurs when $j_{\rm p} \gtrsim j_{\rm f}$ and the jets maintain a more or less constant axis. The jets are inefficient in expelling mass from the equatorial plane and the long-lasting accretion process turns the NS into a BH. 

Although I take the ratio $j_{\rm p}/j_{\rm f}$ to be the main factor that determine the CCSNR morphology, it is definitely not the only one. Other processes might occur, in particular large-scale instabilities during the explosion process. Then there are possibilities of the presence of a binary companion, a circumstellar material into which the ejecta expand, and the interstellar medium. For these, it is expected that opposite structural features, like opposite ears and arcs, will not be equal to each other.   

Although the morphologies of all 14 CCSNRs have been analyzed in the past (see figure captions), this study reports two new results. The first is the classification of CCSNRs to five classes based on jet-shaped morphological features. The second new result is the attribution of the morphological classes to the degree of pre-collapse core rotation as the main (but not sole) factor that determine the morphology class of a CCSNR.  

{ I note that by the same physics by which the jets shape CCSNRs, they can account for non-zero polarization in CCSNe, e.g., as \cite{Nagaoetal2023} report recently. \cite{Nagaoetal2023} find that the explosion asphericity is proportional to the explosion energy and note that jets might account for that. I add here that the JJEM can naturally account for this finding. I take their results to support the JJEM. }

Overall, this study adds some support to the argument that jets, in particular jittering jets (the JJEM), explode most, or even all, CCSNe. The complicated nature of the explosion process and the highly-demanding numerical simulations that are required to simulate the JJEM, force progress to be made in small steps.

% ==========================================
\section*{Acknowledgments}
% ==========================================

I thank Aldana Grichener and Dima Shishkin for helpful discussions and comments. { I thank an anonymous referee for helpful comments. } This research was supported by a grant from the Israel Science Foundation (769/20).

%%%%%%%%%%%%%%%%%%%%%%%%%%%
%\section*{Data availability}
%The data underlying this article will be shared on reasonable request to the corresponding author.  
%%%%%%%%%%%%%%%%%%%%%%%%%%%

% %%%%%%%%%%%%  References %%%%%%%%%%%%%%%%%%%%%

\label{lastpage}

\end{document}